\RequirePackage{tikz}
\documentclass[doublecol]{epl2-fixed}
\usepackage{amsmath,graphicx}

\usepackage{ amssymb }
\usepackage{ wasysym }

\usepackage{tikz} 
\usepackage{nicefrac}
\usepackage{txfonts}
\usepackage{ulem}
\author{Fabian M.~Schaller,\inst{1,2,*} Sebastian C. Kapfer,\inst{1,3} James E.~Hilton,\inst{5} Paul W.~Cleary,\inst{5} Klaus Mecke,\inst{1} Cristiano De Michele,\inst{6} Tanja Schilling,\inst{7} Mohammad Saadatfar,\inst{4} Matthias Schr\"oter,\inst{2} Gary W. Delaney,\inst{5,*} Gerd E.~Schr\"oder-Turk\inst{8,1,*}} 
\shortauthor{F.\ M.\ Schaller \etal}

\institute{
  \inst{1} Institut f\"ur theoretische Physik , Friedrich-Alexander-Universit\"at Erlangen-N\"urnberg, 91058, Erlangen , Germany \\
  \inst{2} Max Planck Institute for Dynamics and Self-Organization (MPIDS), 37077 Goettingen, Germany \\
  \inst{3} Laboratoire de Physique Statistique, Ecole Normale Sup\'{e}rieure, UPMC, CNRS, 75231 Paris Cedex 05, France\\
  \inst{4} Applied Maths, RSPhysSE, The Australian National University, Canberra, ACT 0200, Australia\\
  \inst{5} CSIRO Computational Informatics, Private Bag 33, Clayton South, Victoria, 3168, Australia\\
  \inst{6} Dipartimento di Fisica, "Sapienza" Universit\`a di Roma, P.le A.\ Moro 2, 00185 Roma, Italy\\
  \inst{7} Universit{\'e} de Luxembourg, Physics and Materials Research Unit, L-1511 Luxembourg, Luxemburg\\
  \inst{8} Murdoch University, School of Engineering \& IT, Mathematics \& Statistics, Murdoch, WA 6150, Australia\\
  \inst{*} Email: fabian.schaller@physik.uni-erlangen.de, gary.delaney@csiro.au, g.schroeder-turk@murdoch.edu.au \hfill
}


\title{Non-universal Voronoi cell shapes\\in amorphous ellipsoid packings}

\abstract{ 
In particulate systems with short-range interactions, such as granular matter or simple fluids, local structure plays a pivotal role in determining the macroscopic physical properties.
Here, we analyse local structure metrics derived from the Voronoi diagram of configurations of oblate ellipsoids, for various aspect ratios $\alpha$ and global volume fractions $\phi_g$.
We focus on jammed static configurations of frictional ellipsoids, obtained by tomographic imaging and by discrete element method simulations.
In particular, we consider the local packing fraction $\phi_l$, defined as the particle's volume divided by its Voronoi cell volume.
We find that the probability $P(\phi_l)$ for a Voronoi cell to have a given {\it local packing fraction} shows the same scaling behaviour as function of $\phi_g$ as observed for random sphere packs.
Surprisingly, this scaling behaviour is further found to be independent of the particle aspect ratio.
By contrast, the typical {\it Voronoi cell shape}, quantified by the Minkowski tensor anisotropy index $\beta=\beta_0^{2,0}$, points towards a significant difference between random packings of spheres and those of oblate ellipsoids.
While the average cell shape $\beta$ of all cells with a given value of $\phi_l$ is very similar in dense and loose jammed sphere packings, the structure of dense and loose ellipsoid packings differs substantially such that this does not hold true.
This non-universality has implications for our understanding of jamming of aspherical particles.
}

\begin{document}

\maketitle 

The universality of many features of disordered packings of spherical beads, with respect to preparation protocols and system parameters, is manifest in various properties, such as the universal value of the random close packing limit \cite{Scott1962} and the universal distributions for contact numbers \cite{Aste2006localglobal}, free volumes \cite{StarrGlotzer2002,AsteDiMatteoSaadatfarSendenSchroeterSwinney:2007} and Voronoi cell shape measures \cite{Schroederturkepl2010,StarrGlotzer2002}.
While ellipsoidal particles \cite{Donev2004-1,Donev2004-2,Baram2012,XiaZhuCaoSunKouWang:2014,Delaney2011,Mailman2009,Zeravcic2009,BauleMakse:2014,BauleMariLinPortalMakse:2013,Wegner2014} and other aspherical particles \cite{BauleMakse:2014,BauleMariLinPortalMakse:2013,Delaney2010,Delaney2005,Williams2003,athanassiadis2014,Wegner2014,Baker2010,Haji-Akbari2009,Jaeger2015,Neudecker2013,Smith2014,Torquato2009,Blouwolff2006,Borzsonyi2012b,Fu2012,Kyrylyuk2011,Wouterse2009,Xiao-dan2014} are receiving increasing attention, these questions of universality, including independence of system parameters and preparation protocols, have not been comprehensively addressed yet.
A qualitative difference between ellipsoid and sphere packings is revealed by the analysis of the Voronoi diagram of ellipsoid packings from various experimental and simulated origins.

\begin{figure}[t]
\centering
\includegraphics[width=0.30\columnwidth]{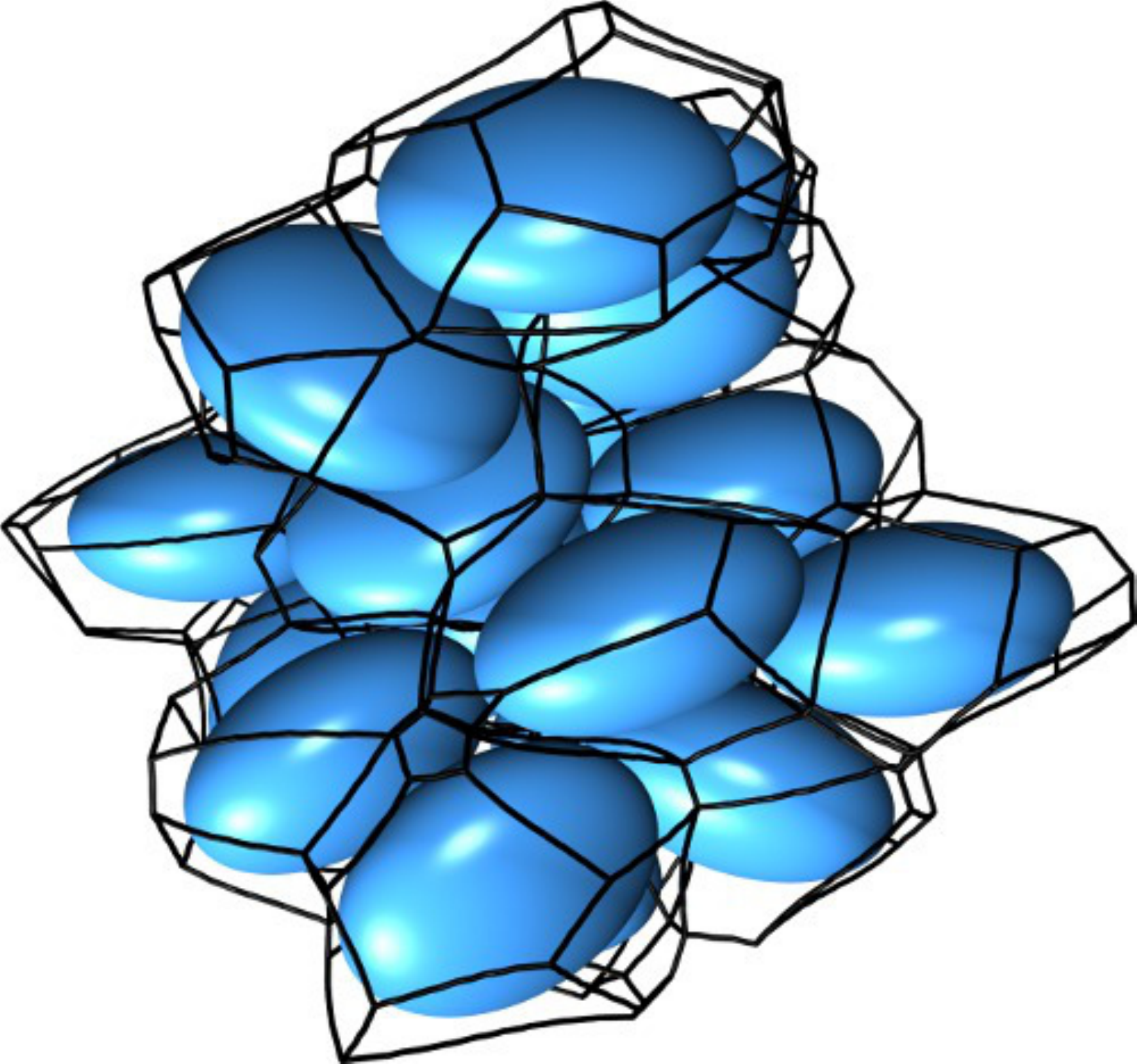}
\includegraphics[width=0.30\columnwidth]{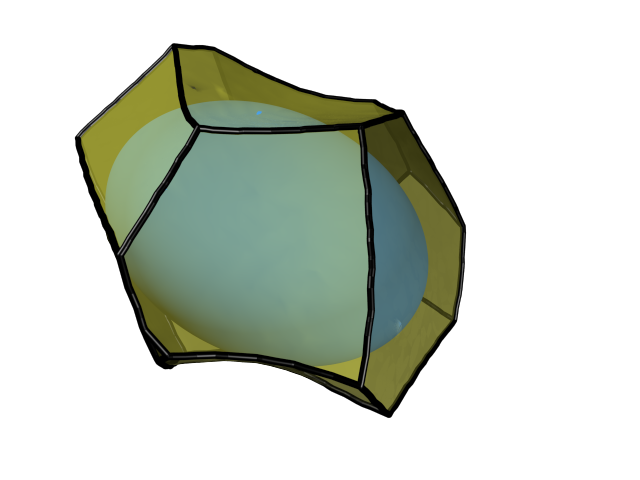}
\includegraphics[width=0.30\columnwidth]{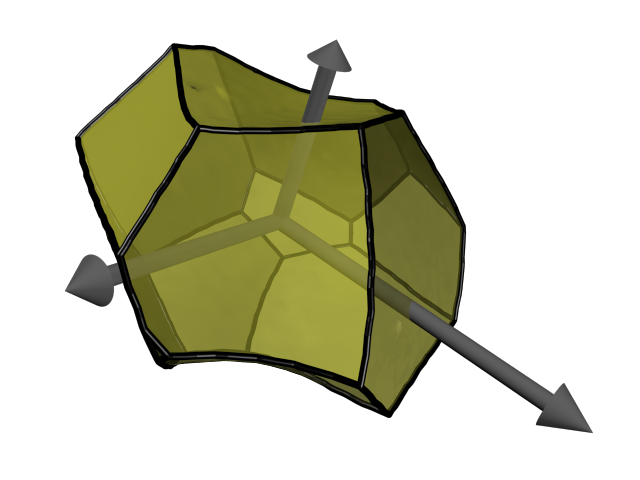}
\caption{\label{f:EllipsoidVoronoiEdges} The Set Voronoi Diagram of ellipsoidal particles has, in general, non-convex cells and curved facets and edges.
            ({\it left}) Subset of an ellipsoid packing with Set Voronoi cells.
            ({\it centre}) Single particle with its cell.
            ({\it right}) Eigensystem of the cell.}
\end{figure}

\begin{table*}
\centering
\begin{tabular}{l l l l l l }
    Data set                       & symbol               & jammed & gravity & friction ($\mu$)    & \#particles \\ \hline
    Experiment                     & $\square$            & yes    & yes     & 0.38 - 0.75      & $\approx$ 5000        \\ 
    Discrete Element Method (DEM)  & $\blacktriangle$     & yes    & yes     & no           & 9323        \\
    Discrete Element Method (DEM)  & $\Diamondblack$      & yes    & yes     & 0.01 - 1     & 9323        \\
    Discrete Element Method (DEM)  & $\blacktriangledown$ & yes    & yes     & 1000         & 9323        \\
    Molecular Dynamics (MD)        & $\circ$              & no     & no      & no           & 512         \\
    Monte Carlo (MC)               & $\bullet$            & no     & no      & no           & 512         \\
    Expansion technique            & $$\LEFTcircle$$      & yes    & no      & no           & 1025        \\
\end{tabular}
\caption{Properties of the data sets analysed in the present work. See main text for details.
\label{tab1}
    }
\end{table*}

\section{Preparation protocols}
The experimental datasets (represented by the symbol~$\blacksquare$) comprise packings prepared by different protocols (fluidised beds, different funnels, grids, pouring particles, etc.) and compaction by vertical tapping.
The ellipsoids have half-axes $a:c:c$ with $a\le c$ and the aspect ratio is defined as $\alpha = a/c$.
These datasets are the same as those used in ref.\ \cite{SchallerContactNumbers2014}, comprising in total 73 datasets of jammed oblate ellipsoids of 5 different aspect ratios $\alpha$ and two different particle types (rough 3D printed particles with $\alpha= 0.4, 0.6, 0.8, 1.0$ and smoother pharmaceutical placebo pills with $\alpha= 0.59$); the larger half-axis is c = 3mm (smallest particles) to 4mm (largest particles).
The standard deviation of the particle volumes is \mbox{2-3 \%}.
The packings were imaged by X-ray tomography; image processing \cite{SchallerTomoMethodSydney13} was used to extract particle center coordinates and orientations.
The packings consisted of approx.\ 5000 particles, of which 600-900 were sufficiently far from cylinder walls to be included in the analysis; for the sake of spatial homogeneity, all packings where radial variations of the packing fraction about the mean exceed 0.66\% were discarded (as in ref.\ \cite{SchallerContactNumbers2014}).   
The particle positions and orientations of all experiments reported here can be downloaded from the Dryad repository \cite{dryad}.

We compare our experimental packings to 120 configurations obtained from discrete element simulations (DEM, \cite{Delaney2011}).
These include datasets of ellipsoids sedimenting into a square box in a viscous fluid under the action of gravity, both of frictionless particles ($\blacktriangle$, RCP) giving highest values of $\phi_g$ and particles with very high friction coefficient ($\blacktriangledown$, Sedimented Loose Packing, SLP) giving looser packings, and intermediate values of friction and viscosity ($\Diamondblack$).
The RCP estimate of the sedimented data is consistent with configurations obtained via a particle expansion method ($\LEFTcircle$) \cite{Delaney2010}.

Reference data for equilibrium configurations of hard-core ellipsoids without gravity in the isotropic fluid phase (Fig.\ \ref{f:anisotropy_global}) was obtained by event-driven molecular dynamics simulations \cite{Michele2010} ($\circ$, MD, the same data sets as in ref. \cite{Michele2007}) and by canonical Monte Carlo simulations ($\bullet$, MC, \cite{Pfleiderer2008}).

\section{Local structure metrics from the Set Voronoi diagram}

For aspherical particles the {\em Set Voronoi diagram} \cite{SchallerSetVoro2013,LuchnikovMedvedevOgerTroadec:1999,LuchnikovGavrilovaMedvedevVoloshin:2002} provides a natural definition of the partition of space into $N$ cells, each containing one of the $N$ particles.
The Voronoi cell of particle $i$ is the compact set of points closer to particle $i$ than to any other particle.
The distance from a point in space to a particle is measured as the Euclidean distance to the nearest point on the bounding surface of the particle.
This is in contrast to the conventional Voronoi diagram where the distance is measured with respect to the particle centre.
Facets of the Set Voronoi diagram are in general curved and cells are non-convex, see Fig.~\ref{f:EllipsoidVoronoiEdges}.
For monodisperse spheres, the Set Voronoi diagram, henceforth simply referred to as the 'Voronoi diagram',  reduces to the conventional Voronoi diagram.
  
The local packing fraction of particle $i$ is defined as \mbox{$(\phi_l)_i=\nu_e/\nu_i$} where $\nu_e=4\pi a c^2/3$ is the volume of the particle and $\nu_i$ the volume of the Voronoi cell $K_i$ containing particle $i$.
We characterise the shape of the Voronoi cell $K_i$ by its volume moment tensor $W_0^{2,0}=\int_{K_i} \mathbf{x} \otimes \mathbf{x} \, dv$ where $\mathbf{x}$ is the position vector relative to the center of mass $\mathbf{c}_i$ of $K$. Similar to the tensor of inertia, this tensor captures the distribution of mass; the notation $W_0^{2,0}$ derives from the theory of Minkowski tensors and integral geometry \cite{AdvMat11,NJP13,Mickel2013}.
The three eigenvalues of this tensor are $\mu^{min}_i \le \mu^{mid}_i \le \mu^{max}_i$. 
The ratio of minimal to maximal eigenvalue $\beta_i=\nicefrac{\mu^{min}_i}{\mu^{max}_i}\in (0,1]$ is an indicator of the shape anisotropy of the Voronoi cell $K$ of particle $i$.
Small values of $\beta_i$ indicate elongated (anisotropic) cells.
Note the difference to measures of asphericity \cite{StarrGlotzer2002} that quantify deviations from a spherical shape; the measure $\beta_i$ is 1 (and $K$ said to be {\it isotropic}) for any shape that has statistically identical mass distribution in any set of three orthogonal directions; this includes the sphere, but also regular polyhedra and the FCC, BCC and HCP Voronoi cells \cite{Kapfer2012}.

\section{Probability distribution of Voronoi cell volumes}

The distribution of the Voronoi cell volumes of sphere packs has been studied in the context of granular materials \cite{AsteDiMatteoSaadatfarSendenSchroeterSwinney:2007,SchallerContactNumbers2014}, super-cooled liquids \cite{StarrGlotzer2002}, and also with respect to granular entropy and the Edwards ensemble \cite{Edwards1989,anikeenko2008,Zhao2014,Song2008,Picaciamarra2012,Bi2014,Puckett2013,Paillusson2012,Wu2015,Asenjo2014,Kumaran2005}.
Aste {\em et al.} \cite{AsteDiMatteoSaadatfarSendenSchroeterSwinney:2007} have shown that in random jammed sphere packings below the RCP limit, the distribution of Voronoi volumes is universal and independent of the preparation protocol.
Starr {\em et al.} \cite{StarrGlotzer2002} have obtained a similar result for super-cooled liquids.
In both cases, the distributions of Voronoi cell volumes collapse when plotted as function of $(\nu-\langle\nu\rangle)/\sigma$ where $\sigma$ is the standard deviation of the distribution $P(\nu)$ and $\langle \nu\rangle$ its average.
Aste {\em et al.} \cite{AsteDiMatteoSaadatfarSendenSchroeterSwinney:2007} proposed a derivation for a scaling $P(\nicefrac{(\nu-\nu_{min})}{(\langle \nu\rangle-\nu_{min})})$ where $\nu_{min}$ is the smallest possible Voronoi cell of equal-sized spheres and $\langle \nu \rangle=\sum_{i=1}^{N} \nu_i/N$ the average over all Voronoi cells.

\begin{figure}
\centering
\includegraphics[width=\columnwidth]{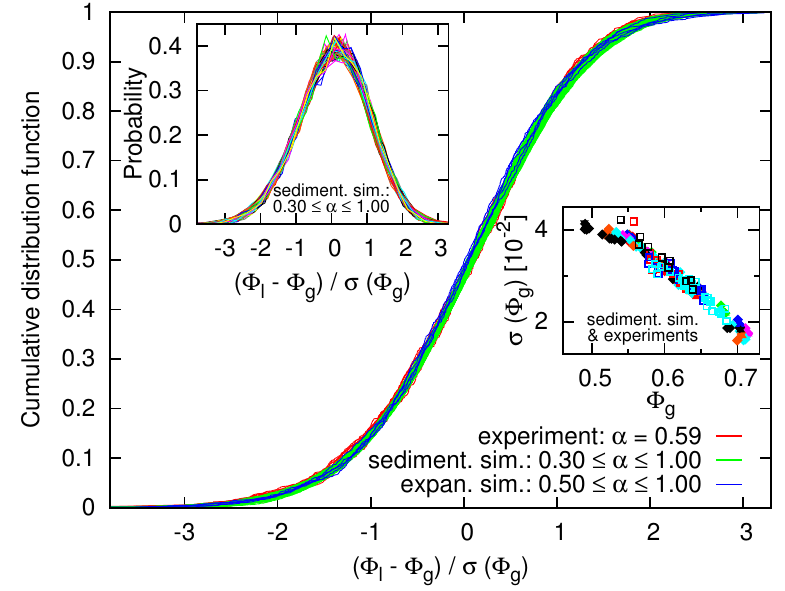}
\caption{\label{f:scale} 
  Probability distribution of local packing fractions $\phi_l$ of jammed static ellipsoid packings (see Table \ref{tab1}).
  Both the cumulative distribution function (main plot) and the probability density (left inset), comprising data sets between SLP and RCP with different aspect ratios $0.3 \le \alpha \le 1.0$ and friction coefficients $0 \le \mu \le 1000$, support the universality of the probability distribution of Set Voronoi cell volumes.
  The right inset shows the collapse of the standard deviations.
}
\end{figure}

Figure \ref{f:scale} demonstrates that this universality is not restricted to sphere packings, but holds more generally for jammed ellipsoid packings:
To the resolution of our data, the functional form of the distribution depends neither on the global packing fraction $\phi_g$ nor on the particle aspect ratio $\alpha$.
The probability for a Voronoi cell in a jammed configuration with global packing fraction $\phi_g$ to have local packing fraction $\phi_l$ is written as $P(\phi_l\mid\phi_g)$.
When plotted as $\sigma P(\phi_l\mid\phi_g)$ vs.\ $\nicefrac{(\phi_l-\langle \phi_l\rangle)}{\sigma}$, it is invariant for all values of $\phi_g$ and $\alpha$, see also Ref.~\cite{SchallerContactNumbers2014}.
This plot shows good agreement between the experimental packings and jammed packings from simulations across the range of accessible packing fractions (those between SLP and RCP, see Fig.\ \ref{f:anisotropy_global}) and aspect ratios $0.3\le \alpha \le 1$.
By contrast, data from equilibrium configurations does not rescale to the same curve.

\section{The shape and anisotropy of the typical Voronoi cell (global averages)}

\begin{figure} 
\centering
\includegraphics[width=\columnwidth]{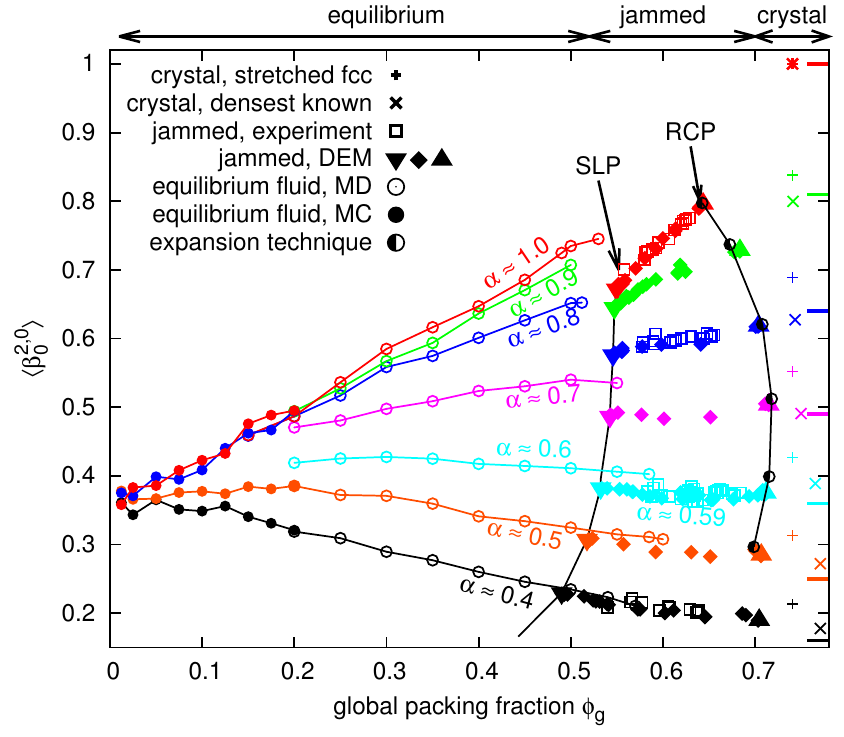}
\caption{\label{f:anisotropy_global}
            Average anisotropy $\langle\beta\rangle$ of the Set Voronoi cells of the ellipsoids as function of $\phi_g$ for equilibrium ellipsoid configurations, static jammed ellipsoid and crystal packings.
            Dashes on the right-hand vertical axis mark the anisotropies of the particles themselves, i.e.\ $\beta$ evaluated for a particle rather than its Voronoi cell.}
\end{figure}

Treating the Voronoi cell volume, or equivalently the local packing fraction $\phi_l$, as the leading term of a shape description of the Voronoi cells, we now proceed to higher-order terms.
While other scalar quantities, such as surface area, integrated curvatures or asphericities may contain signatures of such higher-order terms, we here use the tensorial shape measure $W_0^{2,0}$, similar to the tensor of inertia, and its eigenvalue ratio $\beta$ to quantify the elongation of a cell.

Figure \ref{f:anisotropy_global} shows the average Voronoi cell shape anisotropy, quantified by $\langle \beta \rangle = \sum_{i=1}^N \beta(K_i)/N$, as function of global packing fraction $\phi_g$.
Data is for equilibrium ellipsoid fluids, experiments and simulations of jammed random ellipsoid packings and for two dense crystalline configurations (the {\em stretched fcc} obtained by scaling the $x$ coordinate of the fcc sphere packing, and the densest known structures discussed by Donev {\it et al.}\ \cite{Donev2004-2}). 

For equilibrium fluids in the limit of vanishing density \mbox{$\phi_g\rightarrow 0$}, where the typical distance between particles is large compared to the particle size, the Voronoi cell shape is independent of the particle shape.
Consequently, the shape anisotropy corresponds to the value $\beta\approx 0.37$ of the Poisson point process \cite{Kapfer2010}.
For denser equilibrium fluids \cite{Michele2007, Pfleiderer2008} the trend of the Voronoi shape anisotropy measure $\langle \beta\rangle$ can be understood by realising that the shape of the Voronoi cells becomes more similar to the shape of the particle itself when $\phi_g$ increases
(see dashes on the right hand vertical axis in Fig.\ \ref{f:anisotropy_global}, evaluated for an ellipsoidal particle itself, rather than its Voronoi cell, the ratio is $\beta=(\nicefrac{a}{c})^2=\alpha^2$, see appendix of ref.\ \cite{NJP13}).
When $\alpha$ is small, the curve $\langle \beta\rangle(\phi_g)$ hence decreases, while for larger $\alpha$, $\langle \beta\rangle(\phi_g)$ increases with $\phi_g$. 

For the jammed packings, between SLP and RCP, our results for spheres ($\alpha=1$) are in quantitative agreement with previously published data \cite{Schroederturkepl2010}, with the cells becoming less elongated upon compaction, i.e.\ $\beta$ increases with increasing $\phi_g$.
For ellipsoids with smaller value of $\alpha$, the slope of $\beta(\phi_g)$ becomes smaller and eventually even adopts slightly negative values for small $\alpha < 0.60$.
There is an excellent agreement between the experimenal packings ($\square$, with different preparation protocols) and the numerical data points from DEM simulations ($\blacktriangledown$, $\Diamondblack$, $\blacktriangle$).
As previously found \cite{NJP13}, sphere configurations exhibit a gap in shape anisotropy between the densest equilibrium configuration and the loosest jammed states.
For ellipsoids, this discontinuity shrinks as the particle's aspect ratio decreases.

\section{The shape and anisotropy of the typical Voronoi cell of a given size (local analysis)}

\begin{figure}[t]
\centering
\onefigure[width=0.95\columnwidth]{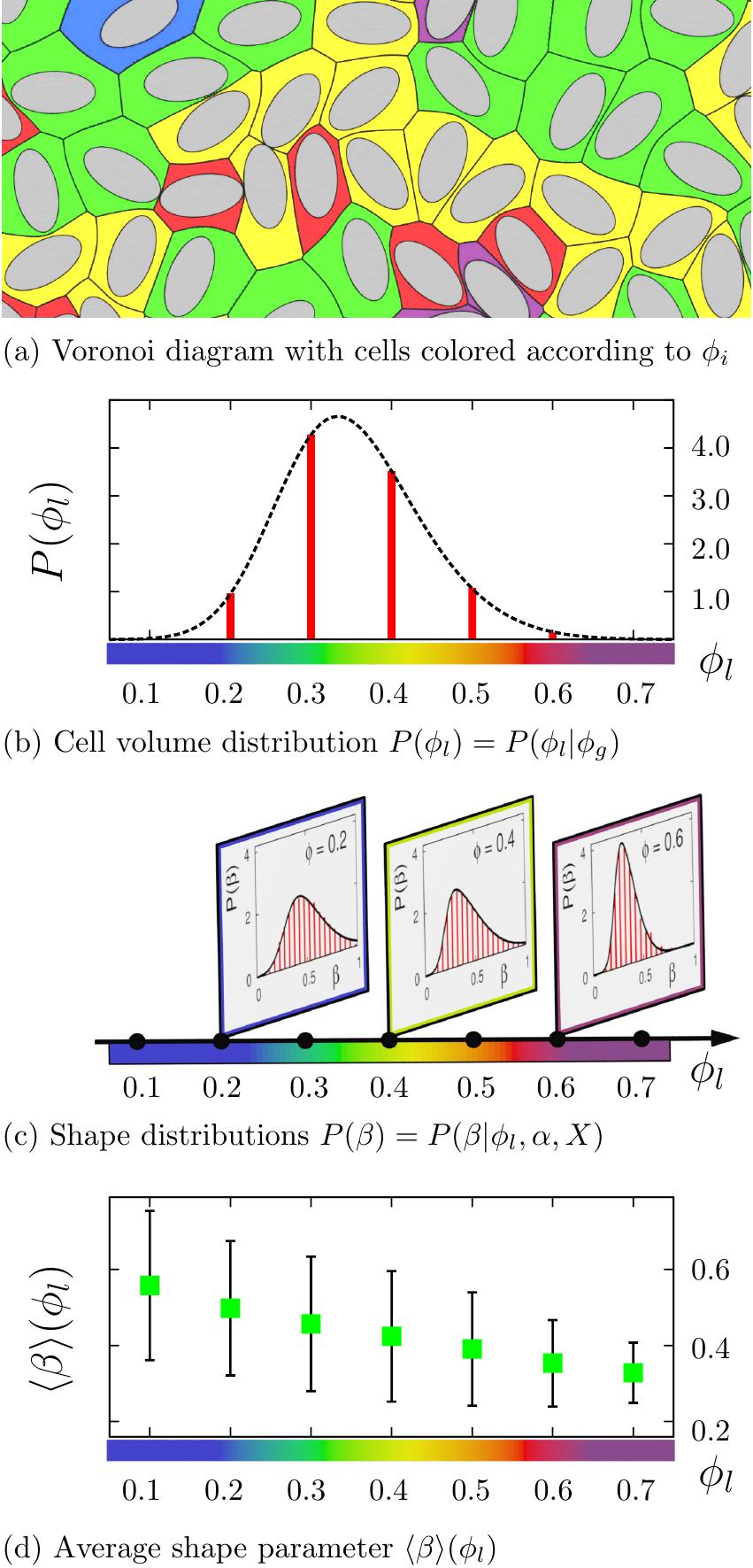}
\caption{\label{fig:density-resolved-analysis-concept}
    The concept of the density-resolved local structure analysis, here illustrated w.r.t.\ $\beta$ for a 2D system.
    Data in this plot are generated from 100 configurations of Monte Carlo simulations of a planar equilibrium hard-core ellipse fluid, of 1944 particles with $\alpha = 0.50$ and $\phi_g=0.34$.
    Error bars in (d) represent the standard deviation of the distributions in (c).}
\end{figure}

The packing fraction $\phi_g$ represents a useful global parameter, which is easily accessible in experiments.
However, there is no conceivable mechanism by which a locally defined quantity, such as the Voronoi cell shape or also the contact number, can depend {\em directly} (i.e.\ by an immediate causal relation) on $\phi_g$; a particle cannot sense the global packing fraction.
That said, in packings with sufficient spatial homogeneity, correlations between averages of the local shape metrics and the global packing fraction are evidently possible, and commonly observed.
Specifically, the study of the average contact number $Z$ as a function of $\phi_g$ is a foundation of the jamming paradigm \cite{OHern2003, SchallerContactNumbers2014}.

Here, we use a local density-resolved analysis based on the idea that the physical mechanisms underlying granular matter occur at the particle scale.
This idea was applied to contact numbers in Ref.~\cite{SchallerContactNumbers2014} and is applied here to Voronoi cell shapes.
Observed correlations between a {\it local} structure metric and the local packing fraction $\phi_l$ are hence more likely to yield physical insight than those with the {\it global} average $\phi_g$.
A similar approach has been used for the analysis of sphere packings \cite{Aste2006localglobal,Schroederturkepl2010}.

\begin{figure}[t]
\centering
\includegraphics[width=\columnwidth]{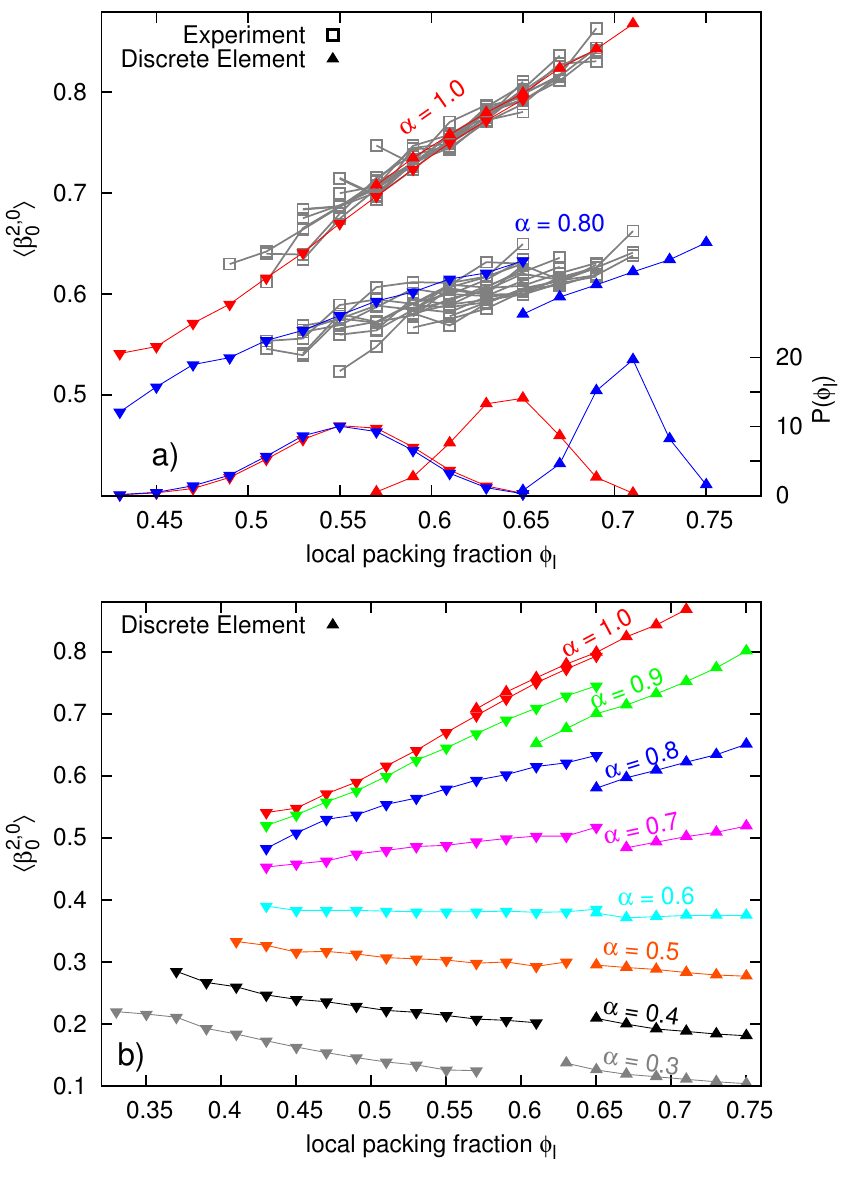}
\caption{\label{f:anisotropy_local}
    Relationship between local packing fraction $\phi_l$ and the Set Voronoi cell anisotropy $\beta$.
    Each curve represents data for a fixed value of $\phi_g$, and is averaged over three independent realisations.
    {\bf a)}~Local analysis for spheres ($\alpha=1$) and slightly oblate ellipsoids ($\alpha=0.8$).
    Gray data points represent experimental configurations, for all available values of $\phi_g$ between RLP and RCP, and the red and blue curves show DEM simulation results for SLP and RCP configurations.
    Curves on the bottom (using the right hand abscissa) show $P(\phi_l)$ for $\alpha=0.8$.
    {\bf b)} Local Voronoi cell anisotropy $\langle \beta\rangle (\phi_l)$ for the DEM estimates for SLP and RCP packings.
}
\end{figure}

Figure \ref{fig:density-resolved-analysis-concept} illustrates the concept of the local density-resolved analysis.
Particles are grouped by their local packing fraction $(\phi_l)_i$, i.e.\ into sets $\mathcal{S}({\phi_l})$ composed of all particles $i$ with $\phi_l-\nicefrac{\Delta}{2} \le (\phi_l)_i < \phi_l+\nicefrac{\Delta}{2}$ for $\phi_l=\Delta, 2\Delta, 3\Delta,\dots$ with a small interval $\Delta$ ($\Delta=0.1$ in Fig.\ \ref{fig:density-resolved-analysis-concept},  $\Delta=0.02$ in Fig.\ \ref{f:anisotropy_local}).
We define the function $P(\beta\!\mid \phi_l, \alpha, X)$, which is the probability distribution of the shape measures $\beta$, restricted to the cells in $\mathcal{S}(\phi_l)$, i.e.\ to those with local packing fraction $\phi_l$.
The unknown parameters $X$ capture influences from the packing protocol, friction etc.
As a result, the $X$ can correlate with $\phi_g$ even though there need not be a causal dependency of the $X$ on $\phi_g$.
The average $\langle \beta\rangle(\phi_l, \alpha, X) = \int \beta \, P(\beta\!\mid\phi_l, \alpha, X) \, d\beta$ over all cells in $\mathcal{S}({\phi_l})$ provides information on how local structure changes depending on local packing fraction $\phi_l$.
In general, $\langle \beta\rangle$ also depends on the aspect ratio $\alpha$ and the unknown parameters $X$.

Figure \ref{f:anisotropy_local} shows the result of this local structure analysis of $\beta$ of jammed ellipsoid configurations.
The key result is the following difference between sphere and ellipsoid packings: in sphere packings, the average shape of the Voronoi cells of a {\em given} local packing fraction $\phi_l$ is, as far as it is captured by the anisotropy index $\beta$,
almost identical in dense and loose packings.
This is evidenced by the near-collapse of the curves $\langle \beta \rangle (\phi_l, \alpha = 1, X)$ for packings of different global packing fraction.
$\langle \beta \rangle$ is a function of $\phi_l$ only, but is largely independent of the unknown parameters $X$, the packing protocol and the particle friction.

In ellipsoid packings, illustrated for $\alpha=0.8$ in Fig.\ \ref{f:anisotropy_local}(a), the curves for different $\phi_g$ do not collapse.
The average $\langle \beta \rangle (\phi_l, \alpha, X)$ depends on both $\alpha$ and $X$.
This indicates that packings with low and high $\phi_g$ exhibit differences in their local structures controlled by $\alpha$ and $X$.
Figure \ref{f:anisotropy_local}(b) demonstrates the validity of this result for other aspect ratios.
It is evident that the densest and loosest simulated packings have different structures, except around $\alpha = 1.0$ (spheres) and $\alpha = 0.6$ (close to the densest random ellipsoid packing).

\section{Discussion and Conclusion}

We have analysed the Voronoi diagram of oblate ellipsoid packings, establishing which aspects of the Voronoi diagram are universal, i.e.\ independent of preparation protocol and particle aspect ratio $\alpha$ and further parameters $X$, and which ones are not.
Considering the geometric nature of this packing problem, these results have ramifications for our understanding of jammed systems and disordered solids. 

The fact that ellipsoidal particles produce denser random packings than spheres is well established, with quantitative agreement between different studies of the value of the 'random close packing' limit $\phi_{RCP}(\alpha)$ as function of aspect ratio for oblate ellipsoids \cite{Donev2004-1,Delaney2010,Delaney2011}, see the curve labelled RCP in Fig.\ \ref{f:anisotropy_global}.
However, while a mean-field theory for $\phi_{RCP}(\alpha)$ is developing \cite{BauleMariLinPortalMakse:2013}, an intuitive geometric understanding for $\phi_{RCP}(\alpha)$ is lacking.
In this regard it is noteworthy that at $\alpha\approx 0.65$, the aspect ratio for which $\phi_{RCP}(\alpha)$ is highest \cite{Donev2004-1,Delaney2010,Delaney2011}, the Voronoi cell shapes are found to be independent of the packing fraction.
As far as captured by $\beta$ of the volume moment tensor $W_0^{2,0}$, the shapes remain approximately constant for all jammed packings, both in the global (Fig. \ref{f:anisotropy_global}) and in the density-resolved analysis (Fig.  \ref{f:anisotropy_local}(b)).

The results of Fig.\ \ref{f:anisotropy_local} emphasise an important distinction between random packings of spherical beads and those of aspherical beads.
The structure of spherical bead packs is universal in the following sense:
on average, the local structure of the typical particle of a given fixed but arbitrary local packing fraction $\phi_l$ is very similar in differently prepared packings, in particular with different $\phi_g$.
This observation, here made w.r.t.\ the Voronoi cell anisotropy of the volume moment tensor, is consistent with similar results for local contact numbers \cite{Aste2006localglobal,SchallerContactNumbers2014}.

It implies that, at least with respect to averages of the volume tensor shape measure, the following interpretation of random jammed sphere packs is feasible.
We consider a pool of local structure motifs for each value of the local packing fraction $\phi_l$, given by the distributions in Fig.\ \ref{fig:density-resolved-analysis-concept}(c).
For spheres (but not for ellipsoids), these pools are universal in the sense that, for a fixed value of $\phi_l$, the same pools can be used to construct packings of various global packing fractions $\phi_g$.
A jammed configuration can then be thought of as the composition of randomly drawn elements from the pools; the probability distribution $P(\phi_l)$ determines the fraction of cells to be drawn from each $\phi_l$-pool.
While this clearly does not represent a constructive approach for the generation of disordered bead packs, it illustrates the universal nature of the sphere packing problem, manifest in the fact that the same pools of structural elements are used for all global packing fractions, just in different proportions.
For ellipsoid packings, this universality breaks down and motifs in the $\phi_l$ pools depend on further parameters $X$ and hence correlate with the global packing fraction.
We speculate that this geometric non-universality is likely to be paralleled by a significantly less universal nature of the random close packing problem in aspherical particles.
To get a deeper insight into the structure formation of non-spherical particles it is important to understand the origin of this non-universality.

Beyond these specific results for ellipsoidal particles, our analysis demonstrates the importance of the correct choice for the relevant parameters for the discussion of local structure metrics in granular matter.
An analysis in terms of the local packing fraction $\phi_l$, which may in principal {\em directly} relate to local physical processes, is more meaningful than the conventional analysis in terms of the global packing fraction $\phi_g$.

\section{Acknowledgments}
We thank Weimer Pharma GmbH for providing placebo pills, and Rolf Schilling for insightful discussions. We acknowledge funding by the German Science Foundation (DFG) through the research group "Geometry and Physics of Spatial Random Systems" under grant SCHR-1148/3-2. We thank Francesco Sciortino for the MD data.
\bibliographystyle{eplbib} 
\bibliography{lib}

\end{document}